\documentclass[prl,twocolumn]{revtex4}

\usepackage{graphicx}
\usepackage{dcolumn}
\usepackage{amsmath}
\usepackage {epsfig}

\begin{document}

\makeatother

\title[short title]{ Cosmology and  Gravitation:\\ Summary Talk at the  XXIV  Brazilian Meeting on Particle  and  Fields}
\author{M.D. Maia  }
\affiliation{ Universidade de Brasilia, Instituto de F\'{\i}sica,\\ 70919-970, Brasilia  DF,
maia@fis.unb.br}
\date{\today}

\begin{abstract}
This is  a   brief  summary with comments on selected   contributions  to
the  Cosmology  and Gravitation  section   at the   $24^{th}$ Brazilian   Meeting on  Particle and  Fields (ENFPC XXIV), held at  Caxambu,  from  September 30 to October 4,  2003. 
\end{abstract}

\maketitle

\section{History and Trends}

The  earliest known  publication on Einstein's  Relativity
 in Brazil is  the little book by Amoroso Costa back in 1922 \cite{Costa}.  After that, a   more technical book  by  Pedro Rache  appeared only  in  1932 \cite{Rache}.
Research papers   properly may  be tracked back  to
the early forties, with the collaboration between  Mario Sch\"onberg and   Chandrashekar in Chicago  \cite{Schonberg}. However, regular and systematic  academic  programs similar to the  ones  we have today,  leading  to  graduate  degrees,  started  in the mid  sixties, with F. Kremmer  in the state of   Paran\'a  and  with   C. Oliveira  at the Centro Brasileiro de Pesquisas Fisicas (CBPF) in Rio de Janeiro.  This  gives us an  estimate of  approximately   40  years  of  development  of  an 
effective   academic research on  cosmology and  gravitation     (CGR)  in Brazil.

During this period  we  have grown to  approximately 150  researchers in that  area,  scattered   through  almost all states  of Brazil. The present meeting  counted   with a  total  of   84  contributions  in CGR  alone,  including two invited  plenary  talks,  three  parallel  presentations,  20 short  oral  communications  and  59  poster  presentations.
 Here  we present  a   sampled analysis  of these contributions,
assembled in   just a few categories.  

There is  a noticeable  tendency  towards   cosmology,  with  53 contributions,  against a  only  31  from  topics in local gravitation, a clear demonstration of  a new trend.  This  is hardly surprising, considering  the present  status  of  cosmology  as  a highly experimental  and interdisciplinary part of  physics, which has provided us  with    a vast and  challenging collection of data  waiting for   explanations.

Therefore, it is also no  surprise that  the   two   invited  plenary talks on CGR  were on the   main experimental results of observational cosmology,   the accelerated  expansion  and  the cosmic background  radiation (CMBR).

The first plenary talk by  A. Lassenby (Cavendish): {\em Cosmic Microwave Background - Recent Results and Their Implications for Cosmology},  analyzed   the  results   on the  CMBR  from the  Wilkinson Microwave Anisotropy Probe (WMAP) experiment released  early this year.

After reviewing the present observational situation,
it was pointed out  that those results confirm  previous  measurements  of  the COBE satellite and more recent balloon and ground-based experiments, but evidenced  a  deficit of fluctuation power on the largest scales, and  also a variation of the spectral index in the primordial perturbation. 
In his own conclusion, ``these  measurements have confirmed the parameters of the current 'concordance model' of the Universe, but with some intriguing pointers towards `new physics' " \cite{Lassenby}.

The  second plenary talk by Jerome Martin (Inst. d'Astrophysique de Paris)  {\em   Inflation 
and Precision Cosmology}, reviewed the  inflationary mechanism   in cosmology  and  its  consequences to   the CMBR  anisotropy  and  formation of large scale structures.  The  predictions of  inflationary  theoretical models  were  compared with  the WMAP data,  showing  an agreement  between  them. The talk  also included an interesting  discussion on the trans-Planckian problem of inflation. That is,  the fact that all the scales of astrophysical relevance today were, at the beginning of inflation, smaller than the Planck length.  Finally,   possible observational signatures, in particular in the CMBR anisotropies, leading to topological  definitions were also discussed \cite{Jerome}.

These two   talks more or less set the  trend for  the  great  majority of  the  discussions on Cosmology during the meeting.
Therefore,   we  start with   cosmology,  moving  towards the  local  aspects  of gravitation.

\subsection{Essences of   Inflation}
A number of possible answers  to  two of the  major open questions in today's  cosmology were presented:  what  caused the past inflation and what causes the current accelerated expansion  of the universe?   

The presently  popular  quintessence model  features a slowly decaying scalar field associated with a phenomenological potential with a small energy  density, of the order of today's value of  the  Hubble parameter $H_{0}\sim 10^{-42}$GeV \cite{Caldwell}. One known   criticism  to this model is that   it is difficult to  conciliate such small repulsive new force  with the present efforts to solve the hierarchy problem for fundamental interactions \cite{Carroll}. 
Therefore,  it is natural to ask  if it is possible to build  a  quintessence model  starting from the phenomenology, for  example using   the  known  data  on the dynamics and thermal history of the universe,  instead of postulating a  potential, \cite{Jackson}.
 
 Along a  similar line,   we could also  look  for   a possible relation between the  primordial  inflationary  period and  today's accelerated expansion. Using  the quintessence  model, it has been  conjectured  that  both  accelerated periods   are described by the same  scalar field,  with a  particle  production  associated with a  non-zero coupling between  matter and the inflaton \cite{Ana}.

On the other hand, particle  creation associated with  inflation
can  be  explained in different ways. For example, it may  be  considered as  a result of     the  gravitational field through
an irreversible process,  or  at the expense of the scalar field together with a phenomenological friction term \cite{Daniele}.

It is  also possible  to suppose that that inflation started at finite temperature, leading to  a  super cooling during the first inflation phase, remaining  cool till reaching the  end of that phase, followed by  a  reheating period  \cite{Rudnei}.

The non-linearity  of  the  decay  of  inflation  as  a cause of the  reheating of the universe  was  described  as a long  time regime dynamical system,  characterizing a turbulent state   with  a  non-linear  transfer of  energy  and 
the consequent  thermalization of the created matter  \cite{Henrique}.

The Chaplygin gas  model proposes  a unified  description of  dark matter and  dark  energy with   a  single   substance  specified  by a  hyperbolic  state  equation $p = -C/\rho$, \cite{Billic}.  The observational  constraints  on such  fluids using data  from the  supernova type I (SNIa) experiments,  the statistics of gravitational lensing and the FR 2b radio galaxies  were  also studied,  showing the best fits for the situations close to the model  composed  by the  cosmological  constant  plus cold dark matter  ($\Lambda CDM$) limit \cite{Makler}.

\subsection{Topology  and Topological Defects}
 The    experimental detection of the topology  of the universe and the  compatibility   of  topological defects  with   experimental  data was  discussed  in several occasions during the meeting.
  In particular the  question  concerning  the  determination of  the  topology  of the universe in a  theory that   is based  exclusively on its local Riemannian metric geometry, as  it is  the case of  General Relativity,  leads  naturally  to  the  question  on how to interpret  any possible  experimental  measurements on the  topology of the universe \cite{Reboucas}. The   finiteness of the universe has been academically discussed for  a long time. Back in  50BC,  Lucretius  \cite{Lucretius} wrote:
\begin{center}
\flushleft{\em  ``...the universe  is not  bounded  in any  direction. If  it were, it  would necessarily  have  a limit  somewhere. But  clearly  a thing  cannot have a limit unless  there  is   something  outside to limit it..."}
\end{center}
\vspace{2mm}
However, in face   of  the  recent  observational data from the  WMAP experiment,  it has been  estimated that   the universe may  be finite  or composed  of  finite  cells. In fact using that data,  it was recently  announced that the topology  of the observable universe  can   be that  to be that of a Poincar\`e  dodecahedron  \cite{Luminet}. The  confirmation of such hypothesis still depends on  the detection of pairs of intersection circles   which  should  necessarily  appear in such topology. This is  expected to  be confirmed   in the  next  planned experiments on the CMBR.

On the other hand,  the same  observations  suggest that the universe is nearly (spatially) flat, with the ratio $\Omega=\rho/\rho_{cr}$ of the total energy density $\rho$ to the critical energy density $\rho_{cr}$ being very close to one. This is compatible with the predictions of standard inflationary models which
also  predicts this ratio to be closer to one by several orders of magnitude. Such analysis  has  been  suggested as  a  way to
distinguish  between two  cosmic topologies as opposed to the usual detection based on  pattern repetition.  
This analysis may decide  which topologies  are potentially detectable but not excluded by observation \cite{Mota}. 

 Since  The Chapligyn gas behaves sometimes  as  attractive   and  sometimes as  repulsive depending on the value of the  parameter $C$, it  has been conjectured  that it must be also associated  with a topological  change in the universe.  However, at  least one  analysis  indicates that  the  delectability  of  the topology of  a  universe  dominated by  such  gas  is  very unlikely \cite{Makler2}.

 Previous   measurements  on the  CMBR have indicated  some  incompatibility with the cosmic  strings  structure. This  was  confirmed  by  the WMAP  experiment as   mentioned  in the  plenary  talk  by  Lassenby.  Nonetheless,   some sophisticated  efforts have been made to  place cosmic  strings back on the track:

In one  example    the cosmic  string structure may be  modified by the  addition of  an  electrical  current,  such  that  it  would  adjust  to the measurements of the   CMBR. This   can be implemented within  the scalar-tensor context, showing that the inclusion of electromagnetic properties induces logarithmic divergences \cite{Andre}.

In addition,  torsion may also be  included in a  current-carrying  cosmic string.  The joint effect of  current and torsion would  change the    gravitational force and  the geodesic  equation of  a test-particle moving around such  ``screwed"  string \cite{Valdir}. 

 In a  different approach to the same problem,  the  cosmic string  was   altered  by the inclusion of  a dislocation. However, in doing so,  an  observer traveling along such string with increasing velocity,  would eventually witness the formation of a time machine. This apparently  undesirable effect
can be fixed by considering a quantized scalar field around the string \cite{Lorenci}.

\subsection{Quantum Cosmology}

  A time independent quantum gravity program  based  on the Wheeler-DeWitt (WDW) equation is   generally known as Quantum cosmology  (QC) \cite{Halliwell}.  Since  in this case time   does not exist,  in one interpretation   the  solutions of  the  WDW  equation can be seen as  being  timeless (or  eternal  so to speak). Nonetheless,  to make sense  with the classical limit of the theory, an appropriate classical time variable must be recovered somehow.  In some  models,  the evolution of  a classical  field attached to the  system  sometimes is  used  as  a  time marker.  However,  generally speaking the recovery of  time in QC remains an  open problem \cite{Brown}.

Quantum Cosmology describes  quantum states of  the  universe  and the probability  transition between these states. Therefore,  it  may  also associate  topological changes in  the corresponding classical universe.  We  could   speak of  a quantum  topological transition in ``quantum topology" associated  with  the quantum tunneling.    Some estimates  show that  such   topological transition between    homogeneous and isotropic  hypersurfaces  with different Riemann curvatures  can    be characterized  by the  Greens' function  of the system  \cite{Nelson1}.

 With the  recovery  of  the notion of  classical time and  with  the notion of  topological transition,  we  may ask what  QC would say  about  the  recent  measurements of the CMBR and of the accelerated  expansion of the universe \cite{Nelson2}. One possible answer  is  through 
 the imposition of  a classical variational principle. Using the Schultz variational formalism, the notion of time can be introduced  so as to obtain  the  FRW  model by a superposition of discrete stationary solutions to the WDW equation. Numerical simulations appear to agree  with results obtained through time-independent perturbation theory. It turns out that these models do not have singularities but they  exhibit quantum tunnelings. Since  this  may  imply in  the loss of  unitarity, the probability transition between  states of the  universe  is not necessarily  conserved, suggesting  that the   ``many-worlds"  interpretation could be applied to  QC \cite{Nivaldo}.

\subsection{Gravitation in Higher and Lower-Dimensions} 

The  four-dimensionality of  space-time  is  an  experimental fact.  At least  that  is  what we  conclude from  Mikowski's  own words  \cite{Minkowski}

"{\em    The views  of  space and time    which  I wish  to lay  before  you have sprung from the soil  of  experimental physics and therein lies  their strength.  They are  radical... " }
 
Of  course, he  was referring to  the Michelson-Morley  experiment and   at that time nothing else would  indicate  that we need more than  four  dimensions. However,  theory, experimental  high energy physics  and astrophysics  have  evolved  enormously  since  that  period, suggesting that   we may need more than  four  dimensions  to describe  space-time.

In fact,  gauge field theory  is consistent with  the addition of a  set of parameters  associated  with  a symmetry  group which exists together with the  Poincar\`e symmetry. This  characterizes an  ``internal space" which is  observable  through  high energy physics  experiments. Therefore, today we could rightfully  repeat  Minkowski's  words, stating   that  ``higher dimensional physics  also  sprigs  from the  soil  of  experimental physics". The difficulty  lies 
in finding the  correct  way  to  combine  all  degrees  of  freedom in a  single space  with dimension greater  than  four.  
  
One of  the  most recent proposal in that  direction is the  scheme  of  ``large extra  dimensions", also known as  the brane-world program, where  the four-dimensional space-time is a  submanifold  embedded in a  higher dimensional  space \cite{ADD,RS}. The extra dimensions can be as large  as a micrometer and  there  are  serious  experiments under way to verify Newton's gravity at  short distances \cite{Mostepanenko}. Additional
experiments are being planned for  2007, aiming to 
probe  the  extra  dimensions  at the TeV  scale of  energies \cite{Dimopolous}. 

The  accelerated expansion of the universe as  an  observed  fact has  been also  understood as  an  experimental  evidence  for  the existence of  large extra dimensions.  This interpretation is  possible  because  large extra dimensions  suggest a
 modification  of    Friedmann's  equations by an additional  term. In most  models  this 
 term can is proportional  to the square of the energy density \cite{MaiaFr}, which    does not agree  with the  formation of  light  elements  from the big-bang  nucleosynthesis,  suggesting a number of rather  creative  fixes. 

One of these repairs  consists  in  modifying the    energy  scale of the nucleosynthesis. Admitting that the fundamental scale  is   1 TeV  and   using  a  semi-analytic  procedure, the primordial Helium abundance  appears to be 
 in agreement  with  the  observations \cite{Fabris}.

 A  higher  dimensional  Minkowski space  has  also been considered  as  the  host space (the bulk) in which the  brane-world moves as  an  extended object \cite{DGP}.  In this case,  the motion of  the brane-world emits  gravitational waves  in the bulk,  which may become  a  source of  gravitational  waves itself. In the case of    bulk waves produced by the linear perturbations of  the bulk, we obtain besides  spin two fields, other gravity related  objects whose nature depends on the  number of  extra  dimensions.  Therefore, the higher-dimensional gravitational  dynamics  produces  a difference relative to  the usual  four-dimensional one when   counting  the  spin degrees  of  freedom \cite{Damour}.

 The dependence of  the dynamics  of an object  on   the dimension of the  space-time also appears  in  
different contexts.  For  example,  it has been  shown that the emission rate for a source with constant proper acceleration in a  D-dimensional  Minkowski space-time,
minimally coupled to a massless scalar field, is equal to the emission plus absorption rates when the static source is in a thermal state.  The response rate is generally proportional to a certain power of the source's constant proper acceleration, which depends on the dimension $D$ of the space-time \cite{Crispino}.

The  Aharonov-Bohm  effect indicates  that the
 electromagnetic  potential  is  an  observable. By  extension  a  gauge  potential is  also  taken to be  an  observable\cite{Moriyasu}.
Therefore, looking way ahead at  a quantum version of  brane-worlds, as realized through the Ashtekar-Smolin loop  quantum gravity program \cite{Smolin},  we may  ask  if  the   potentials derived  from  the  holonomy of a triad in a brane  would  also be an observable  in the sense of the gravitational analogue to the Aharonov-Bohm effect.  An estimate of such effect  for  the  cases  the Schwarzschild  and  AdS-Schwarzschild space-times,  in the five-dimensional Randall-Sundrum scenario,   was   presented \cite{Furtado}.

Since the  brane-world  program  sits  at the  interface  between  gravitation and  field  theory, some of  the contributions on brane-words and applications to cosmology also appeared in the field  theory sections of  the meeting,  where   questions  relative to  the presence of massive  Kaluza-Klein modes  were discussed.  In one talk   massive gravitons are  considered in  thick branes regarded as  domain walls  associated with  tachyon potentials. In the application to cosmology, these  potentials  can be  selected so  as to  describe some form of  dark energy.  In particular a  tachyon matter with  negative pressure is  suggested  as a candidate for quintessence \cite{Bazeia}. In another  talk, the experimental  constraints  to the brane-world program, associated  with the linear gravitational waves emitted  by  the bulk and the  presence of  massive  Kaluza-Klein graviton modes was  discussed  \cite{Durrer,Maia1}.

\vspace{3mm}

Quite on the opposite direction,  there were a number of  presentations  on  lower    dimensional gravity. 
For example, a perfect fluid with a kinematic self-similarity has been  considered  in 2 + 1  dimensions with circular symmetry. In this case,  some solutions of Einstein's  equations represent gravitational collapse  in which the  final state  can be either a black hole or a null singularity \cite{Miguelote}. 

A  study on  inflationary  cosmology in 2+1  dimensions, including  dark energy  solutions,  show that
matter and or radiation  would interact with the  inflaton  field through particle production  during the inflationary regime. In the present era,  the
interaction of matter and  a dark energy  component  represented by the scalar field produce  a  positive acceleration   as  a function of   the total pressure of the  constituents \cite{Carina}.

A similar  idea can also be   applied to  the case of 1+1 dimensions, with a  scalar field  representing  the dark energy,
where  a  comparative analysis is made for both the Jackiw-Teitelboim and the dilatonic models in arbitrary gauges \cite{Cristmann}.

\subsection{Gravitational Wave  Detection}
The  experimental  program of  gravitational wave detection in Brazil started in the early  nineties with the ``Graviton  Project".  Later on
 in 2000 it led  to  the Schenberg project for the construction of  a  smaller spherical antenna,  which consists of the  current experiment,  mostly developed at the  National  Institute for  Space Research (INPE).

 The spherical   detectors  with  $60$cm  diameter  have a   resonance frequency of $3.8$Hz  which is  greater than the  laser interferometry antennas. The 1.15 Ton copper-aluminum sphere and its 0.5 Ton vibration isolation system has been  cooled down to 2 K for a first cryogenic test. 
 
  An initial test run has been  already made and the   first operational run  at 4-5 K,  with resonant frequency at  around 3.2 kHz, is expected to  operate by  mid 2004,  with  nine   transducers  \cite{Odylio}.

 A    simulation of  the   detector was made with the  wave generated by  two inspiralling neutron stars which  eventually collide into a  black hole. 
The signal-to-noise ratio was   calculated as a function of frequency for the simulated signals,  showing  that
the sphere is  sensitive enough to detect that kind of signal up to a distance of  18kpc  \cite{Odylio1}

The development and  construction of the   transducers represents  an 
additional engineering  task, which  has  benefited from the experience from previously   existing   antennas. The  present development  uses  a modification  of  the transducer  used for the gravitational wave bar antenna Niobe, making it  easier
 to  achieve  the standard quantum limit of sensitivity  \cite{Frajuca}.

\subsection{Analog Models}

The  idea behind  analog models consists  in  getting    insights on problems  defined   in a given  branch of  physics  by comparison  with an analog   problem in
another area of  physics or, more generally,  of  science \cite{Novello}.
In the particular case  of  gravitation, one well known  example is  given by  the thermodynamics of    black holes, where Hawking's  radiation  is  regarded  as  the analogous to a thermal  radiation   \cite{Lorenci1}.

In another analogy with  optics, Fresnel's  equation can  be   written  in a covariant  form  aiming to an  analogy  with  wave propagation  in  curved  space-times. The solutions of  that  covariant  equation are intrinsically obtained, showing that the light rays propagate in eigenmodes  with linear polarization 
\cite{Lorenci2}.

An interesting  example is  the analogy  of  dark energy in  cosmology  with  a  classical fluid 
with negative pressure. It is noted that while in non-relativistic physics negative pressure is associated to cohesive forces,  in relativity  it may  prevent the collapse of a mass distribution \cite{Bedran}.

We  expect  that  these  analogies work in both ways, specially  on fundamental issues. For  example,  we may regard  the geometric description of  Yang-Mills  theory as  analogous  to   Einstein's  gravity.
Although  the latter is  not  a gauge theory, it seems possible to derive an  equivalent gauge theory (see below). The  same  equivalent gauge theory may
offer an interesting  analogy  with  solid  state  physics, perhaps with   some  interesting   geometrical tools.

\subsection{Alternative Formulations}

In the past  years there  has been  a noticeable interest  the development of the so  called  Teleparallel Equivalent of General Relativity.  Basically,  instead  of  using  curvature this formulation of Einstein's  gravitational theory uses  only the torsion. 
The  Riemann  curvature of the space-time vanishes and  everything  else,  including geodesic deviations are   handled  in terms of the torsion tensor.  The complete equivalence  between teleparallel gravity and  Einstein's  gravity  depends  on the  equivalence principle but that   seems to be  undecided \cite{Aldrovandi}. If the total equivalence  holds,  then  this  new  formulation may   not offer  much  at  the classical level,  including   possible explanations for   the  dark matter and  energy problem. 
 
  In the past, equivalent  formulations  of    Einstein's   theory like  the  two-component spinor formulation or  the Newmann-Penrose formalism,  have  offered  some  simplifications  on  particular  issues  such as the  Petrov classification and  on the computer-algebra methods of solving   Einstein's  equations.  Later on the  spinor formalism and conformal invariance led to the development  of  the  conformal  spinor or  ``twistor"  program, which   eventually  led to the concept of spin networks.
 
  It is  possible that the contribution of the teleparallel  program  rests  not  on the classical physics but on quantum gravity.   Indeed, a  major  conceptual problem  when we  want  to place  all fundamental interactions in the  same basket is  that Einstein's gravitation does not behave  as  a  
 gauge theory. However, as it has been reported in the meeting,   teleparallel  gravity appears  to be  a
true  gauge theory   of  gravity  of the   Poincar\`e translational group  $T_{4}$  \cite{Geraldo}.  
In this case    we could in principle quantize  gravity through the $T_{4}$ connection, bearing in mind that  the  $T_4$ group  is not an easy one. Besides  being non-compact, it  has the rather undesirable  property  of  acting nilpotently over the other    symmetries,  possibly  leading to an acute    No-Go problem when  the  $T_{4}$  group is  combined  with  the standard  gauge  groups   \cite{Oraifeartaigh,Colleman,Maia}.

\subsection{ Singularities and  Black Holes}

Most of the  contributions  on  exact  solutions of  Einstein's  equations concern   the addition  of a scalar field and to explore
the its  quantum and  classical  
properties,  notably  in   black-hole physics. 
For  example,    Hawking's radiation  can  be determined by  means of a  massless scalar field. In the particular cases  of   Schwarzschild  and  Schwarzschild de Sitter  space-times,   the response rates at the tree level in terms of an infinite sum of zero-energy field
modes can be calculated in  a closed form in 
de Sitter space-time,  but not in the in the Schwarzschild-de Sitter case \cite{Jorge}.
 
The study  of  properties of  black hole  physics  like  the decay of  its  mass and  the area  spectra  cannot be complete  without the analysis  of  quantum gravity. Although  such theory  does not exist  as  yet, an   estimate of  some of these properties can be obtained  from the  basic  properties  a  given quantum gravity scheme.
 
For example, loop Quantum Gravity (LQG) is  based on the auxiliary variables  program  originally developed by  A. Ashtekar. A connection associated with  the  holonomy group of  closed loops  may be  quantized as  a  $SU(2)$ gauge theory,  inducing   the   quantum   fluctuations of  the  metric geometry,  while keeping intact  the diffeomorphism  invariance  of  the  four  dimensional  theory \cite{Smolin}.  Even  if  one such  theory is  far from complete, it  has been  claimed to
 contribute to the  spectra  of  black hole  area \cite{Barreira}.

In   an  analogous problem  but using a   more  traditional approach to quantum gravity, the Arnowitt-Deser-Misner (ADM)  quantization,  together with the  de Broglie-Bohm interpretation of quantum mechanics,
it is  also possible to evaluate  the mass  spectra in the case of  a  spherically symmetric black hole coupled to a massless scalar field. The quantum states of  the black hole mass
can either increase or decrease with time, depending on the
imposed  initial conditions  \cite{Acacio}.

In another application   a massless  
scalar field  may  also  explain the emergence of singularities. For  example, a  self-similar solution to the Einstein's equations in four-dimensions  can be   obtained, representing the collapse of a spherically symmetric, minimally coupled, massless, scalar field  leading to  a  Schwarzschild  black hole \cite{Takakura}.

 The process  for  extracting  energy from a rotating black-hole
 as  proposed by  R. Penrose many years  ago has been  considered  as  a possible explanation for the   gamma ray bursts  \cite{Rufini}.  These are   short  bursts   of  gamma rays at the range oh  100's  KeV,  lasting  a  few milliseconds to  several hundred  seconds \cite{Piran}.
 However, it was questioned  if the    extracted energy is  sufficient and  if  it can be sustained in face  of  other processes going around the black hole, such as the formation of a  plasma  composed of  charged particle pairs. The basic  argument is that  such
plasma  may  remove   sufficient energy  to prevent  the  generation of  gamma ray  bursts  \cite{Opher}.
It  has been noted also that  the extraction of  energy may   be  followed  by  the  superradiance amplification of  amplitude and frequency of  the emitted wave, which  may result in  an   additional  loss of  energy  \cite{Joras}.

\section{Concluding Remarks}
 
  Our  sampled  analysis of the contributions shows that  after  approximately  forty years,  research on gravitation and  cosmology  in Brazil not only stays on the edge  of most recent developments,  but  also has  innovated and leaded  some  lines  of  research.  The  majority of  contributions  are  on  cosmology,  which  we interpret  as  a  natural trend  resulting  from the  very large amount of  data  indicating  the existence of a present  day accelerated  expansion  of the universe  and  the  more recent  precision  measurements     of  the  CMBR  by  WMAP.

The  topology of  the universe  has been systematically  studied  in Brazil  for more than two decades \cite{Fagundes}. We find it  particularly  relevant that  the meeting  counted with several  contribution on the experimental   detection   of the topology of the cosmos \cite{Reboucas,Mota},  specially because,  just  two  weeks after  the meeting,  it was  announced that  the WMAP results  point to  a specific  topology and that  it  could  be  experimentally confirmed  by  the next planned experiments \cite{Luminet}.

Still on  the  experimental facet  of  the new gravitational physics, we cannot  avoid mentioning the collection  of  reports on the Brazilian  Gravitational wave Project.
 The  development of higher frequency oriented   spherical detectors  of  gravitational waves list as  one  of  our  outstanding efforts \cite{Odylio}. Our  expectative  is  that  in the next meeting  we  will have some  positive  data collected  from these detectors.

On  the  theoretical side, we have  spotted  three  relevant  contributions: First, the possibility  or  not  of  explaining    gamma  ray bursts  by means  of  Penrose's  process  of  extracting  energy  from a black hole  has the merit of  opening a debate
on such relevant  matter \cite{Opher}. Secondly,  
the  interesting description of the universe as   a  dynamical  system    appeals to  a mathematical aspect  in which  Brazil  has given  a  substantial   contribution, namely the theory of  dynamical systems \cite{Henrique}.  Thus,   future joint ventures  on  ``universe dynamics"  are expected and  should  be  stimulated.
Finally,    it has been suggested that the gauge  potential associated with the  $SU(2)$ triad holonomy group  should  have  a  observational  character  through  the  gravitational analogue of the Aharonov-Bohm  effect \cite{Furtado}.

It was  an extremely gratifying experience   to  talk  with 
 most of  the   contributors  during the meeting.
We only  regret that  we were unable to comment on many other  
quality contributions   and  to be  limited  to  select a just a  few topics.


\begin{thebibliography}{99}

\bibitem{Costa} M. Amoroso Costa,  Introdu\c{c}\~ao  \`a Teoria  da Relatividade.  Editora UFRJ,   (in Portuguese) Second  ed. (1995)
\bibitem{Rache}  Pedro D. Rache,  Relatividade e  sua Applica\c{c}\~ao   ao Estudo dos  Phenomenos Physicos Precedidos  dos Elementos Indispensaveis de Mathematica.
  Imprensa Oficial de Minas Gerais, 
Belo Horizonte   (in Portuguese) (1932)
\bibitem{Schonberg}  M. Sch\"onberg  \&S. Chandrasekhar,  Astrophysical  Jour.  vol.96, 161 (1942)
\bibitem{Lassenby}  A.  Lassenby,  Cosmic Microwave Background - Recent Results and Their Implications for Cosmology.  Braz. Jour. Phys.  This issue  (2004)   
\bibitem{Jerome}  J. Martin,  Inflation 
and Precision Cosmology. Braz. Jour. Phys.  This issue  (2004)
\bibitem{Caldwell} R. Caldwell  Braz. J. Phys. vol.30 215,(2000)
\bibitem{Carroll}  S. M. Carroll  astro-ph/0107571
\bibitem{Jackson} J.M.F.Maia, Solu\c{c}\~oes Cosmol\'ogicas com Campos Escalares Acoplados,(in Portuguese),(2003)\\ www.sbf1.if.usp.br/eventos/enfpc/xxiv/   
\bibitem{Ana} Ana Helena Campos et al. Preheating in Quintessence Inflation,(in Portuguese)(2003)\\ www.sbf1.if.usp.br/eventos/enfpc/xxiv/ 
 \bibitem{Daniele}  Daniele, S.M. Alves and G. M. Kremer,
Acceleration field of a Universe  Modeled as  a  Mixture of  Scalar and Matter Fields, (2003)\\ www.sbf1.if.usp.br/eventos/enfpc/xxiv/
(2003) 
\bibitem{Rudnei}  R.O.Ramos, The Prevalence of  Nonisentropic Expansion in Inflation Models,  (2003)\\ 
www.sbf1.if.usp.br/eventos/enfpc/xxiv/  
\bibitem{Henrique} H. Pereira and  I.D. Soares,  Preheating and Turbulence: Echoes  of a  not so Quiet Universe, (2003)\\
www.sbf1.if.usp.br/eventos/enfpc/xxiv/ ,  see  also
I. D. Soares,  Phys. Rev. vol.D60, 121301, (1999) 
\bibitem{Billic} N. Billic et al, The Chaplygin Gas and the Evolution of Dark Matter and Dark Energy in the Universe,
 astro-ph/0307214,  (2003)
\bibitem{Makler} M. Makler et al,  Observational Constraints on Chaplygin Quartessence: Background Results,  astro-ph/0306507, (2003)
\bibitem{Reboucas}  M. J. Rebou\c{c}as, Topologia C\'osmica, Braz. Jour. Phys. This issue  (2004)
\bibitem{Lucretius}   Titus Lucretius Carus,  On the nature of the Universe, circa 50  B.C. Penguin Books, (1951)
\bibitem{Luminet} J. P.  Luminet et al, Nature vol.425, 593, (2003), astro-ph/0310253, (2003)
\bibitem{Mota}  B. Mota  et al,  Constraints on the Detectability of Cosmic Topology from Observational Class. 
Quant.Grav. vol.20, 4837, (2003),  gr-qc/0308063, 
\bibitem{Makler2} M. Makler  et al,  Detectability of  Cosmic Topology in the Chaplygin Model  of dark  Matter  and Dark Energy Unification, (2003)\\
www.sbf1.if.usp.br/eventos/enfpc/xxiv/ 
\bibitem{Andre}  M.E.X Guimaraes and  A.L.N. Oliveira,  Wakes in Dilatonic Current-carrying Cosmic Strings. Phys. Rev. vol.D67, 123514, (2003)
\bibitem{Valdir}  V. B. Bezerra  et al,  Gravitational Field Around a Time-like Current Carrying  Screwed  Cosmic String in Scalar Tensor Theories, hep-th/0306289, 
\bibitem{Lorenci}   V. A. Lorenci and E. S. Moreira Jr.
Spinning Strings Cosmic Dislocations and Chronology Protection, gr-qc/0309122
\bibitem{Halliwell} J.J. Halliwell,  The interpretations of  Quantum Cosmology and the Problem of  Time,  gr-qc/0208018, 
\bibitem{Brown} J. David Brown and James W. York, Jacobi's Action and the  Recovery of  Time in General Relativity,
Jr. Phys. Rev. vol.D40, 3312,  (1989)
\bibitem{Nelson1} N.  Pinto Neto et al, 
 Green's Functions for Topology Change,  gr-qc/0006096
\bibitem{Nelson2} Nelson  P. Neto  et al, 
The Accelerated Expansion of the Universe as  a Quantum Cosmological Effect. Phys. Lett. vol.A315, 36, (2003),
gr-qc/0302112 
\bibitem{Nivaldo}  N. A. Lemos, The Role of  Quantum Cosmology in the Chaotic Dynamics of  Inflation, (2003)\\  www.sbf1.if.usp.br/eventos/enfpc/xxiv/ 
\bibitem{Minkowski} H. Minkowski,  Space and Time,  in  The Principles of  Relativity, H. A. Lorentz and others,  Dover Publications, (1923)
\bibitem{ADD} N. Arkani-Hamed et Al,  Phys. Lett. vol.{B429}, 263 (1998), Phys. Rev. Lett. vol.{84}, 586, (2000)
\bibitem{RS}   L. Randall \& R. Sundrum,  Phys. Rev. Lett. vol.{83}, 3370,(1999);   Phys. Rev. Lett.  
vol.83, 4690 (1999).
\bibitem{Mostepanenko}  R. S. Decca et al,   Improved Tests of Extra-dimensional Physics and Thermal Quantum Field Theory From new Casimir Force Measurements,  hep-ph/0310157 (2003)
\bibitem{Dimopolous} S. Dimopolous and  G. Landsberg,  Black Hole Production  at Future Colliders. SNOWMASS-2001-P321,  (2001) (also in Les Houches (2001)). See also, K. Cheung, Collider Phenomenology  for  Models of Extra Dimensions, hep-ph/0305003, (2003)
\bibitem{MaiaFr}  M. D. Maia et al, On  Friedmann's  Equation in Brane-Worlds, Intl. Jour. Mod. Phys.  vol. A17,  29, (2002)
\bibitem{Fabris}J.C. Fabris, J.A. de Oliveira Marinho, (2003)\\
 www.sbf1.if.usp.br/eventos/enfpc/xxiv/
\bibitem{DGP} G.Dvali, G. Gabedadze, M.Porrati, 4-D Gravity on a Brane in 5-D  Minkowski Space, Phys. Lett. vol.B485, 208,(2000),
 hep-th/0005016 
 \bibitem{Damour}  T. Damour  et al,  hep-th/0212155, (2002)
\bibitem{Crispino}L. C.  Bassalo Crispino and G. E. A. Matsas ,
Radiation Emission  from a  Uniformly Accelerated  Source in  n-dimensional  Minkowski Spacetime, (2003)\\ www.sbf1.if.usp.br/eventos/enfpc/xxiv/
\bibitem{Moriyasu} K. Moriyasu, An  Elementary  Primer  for Gauge  Theory, World Scientific,  (1983).
\bibitem{Smolin}  Lee Smolin, How Far Are We from the Quantum Theory of Gravity?  hep-th/0303185,  (2003)
\bibitem{Furtado} A. M.  de Morais Carvalho et al,  Holonomy Transformation in a d-Brane, (2003)\\  www.sbf1.if.usp.br/eventos/enfpc/xxiv/
\bibitem{Bazeia},  D. Bazeia Filho e  J. R. S. Nascimento
Tachyons and Braneworlds, (2003)\\
www.sbf1.if.usp.br/eventos/enfpc/xxiv/,
hep-th/0306284 
\bibitem{Durrer}  R. Durrer  and  P. Kocian,  Testing Brane Worlds with the  Binary Pulsar, hep-th/0305181, (2003)
\bibitem{Maia1} M.D. Maia, Kaluza-Klein Massive Modes in Brane-worlds, (2003)\\ www.sbf1.if.usp.br/eventos/enfpc/xxiv/ 
\bibitem{Miguelote} A. Y. Miguelote  et al,  Gravitational Collapse of Self-Similar Perfect Fluid in 2 + 1 Gravity,
gr-qc/0304035, (2003)
\bibitem{Carina} C.  M.  Zanetti  and F. P. Devecchi, Regimes Inflacion\'ario e de Energia Escura em uma Cosmologia Viscosa em  2+1-dimens˜oes, (2003)\\  www.sbf1.if.usp.br/eventos/enfpc/xxiv/
\bibitem{Cristmann} M. B. Christmann et al, A  Study on Inflation and Dark Energy Regimes in Dilatonic and Jackiw-Teitelboim Cosmologies, (2003)\\  www.sbf1.if.usp.br/eventos/enfpc/xxiv/
\bibitem{Odylio} O.D. Aguiar et al,
The Gravitational Wave detector ``MARIO SCHENBERG": Status of the Project  Braz. J. Phys. vol.32, 866, (2002);  O. D.  Aguiar et al, First Cryogenic Run of the Sphere and Latest Developments, (2003)\\  www.sbf1.if.usp.br/eventos/enfpc/xxiv/
\bibitem{Odylio1}Response of the Brazilian gravitational wave Detector to Signals From a Black Hole Ringdown. Cesar A. Costa  et al   gr-qc/0309047 (2003)
\bibitem{Frajuca} C. Frajuca  et al,   Transducers for The Schenberg Detector: Addresing  Some  Mechanical  Challenges, (2003)\\
www.sbf1.if.usp.br/eventos/enfpc/xxiv/
\bibitem{Novello} Novello et al,  Artificial Black Holes, World  Scientific  (2002)
\bibitem{Lorenci1} V.A. DeLorenci et al,  On Optical Black Holes in Moving Dieletrics, Phys.Rev. vol.D68, 061502, gr-qc/0210104
\bibitem{Lorenci2} V.A. DeLorenci and R. Klippert,  Electromagnetic Light Rays in Local Dielectrics, (2003)\\  www.sbf1.if.usp.br/eventos/enfpc/xxiv/ 
\bibitem{Bedran}M. L. Bedran,   The Effect of Negative Pressure, (2003)\\
 www.sbf1.if.usp.br/eventos/enfpc/xxiv/ 
\bibitem{Aldrovandi}R. Aldrovandi et al, Gravitation Without  the  Equivalence Principle,  gr-qc/0304106 
\bibitem{Geraldo}  J.G. Pereira,  Selected Topics in Teleparalell Gravity,  Braz. Jour. Phys.  This issue  (2004)   
\bibitem{Oraifeartaigh}  L. O'Raifeartaigh, Phys. Rev. Lett.  vol.14, 475, (1965)
\bibitem{Colleman} S. Coleman and  J.  Madula,  Phys. Rev.  vol.159,  1251, (1967)
\bibitem{Maia} M. D.  Maia, An Inversion to O'Raifeartaigh's Theorem,  Rev. Bras. Fis. vol.11, 429 (1981)
\bibitem{Jorge} J.C.  Rodryguez et al,  Intera\c{c}\~ao  da  Radia\c{c}\~ao de Hawking com Fontes Escalares Est\'aticas nos
Espa\c{c}o-tempos de deSitter e Schwarzschild-deSitter (in portuguese), (2003)\\ www.sbf1.if.usp.br/eventos/enfpc/xxiv/
\bibitem{Barreira} M. M. Horta Barreira, A Semi-Classical Quantiztion for Loop Quantum Gravity, (2003)\\ 
 www.sbf1.if.usp.br/eventos/enfpc/xxiv/
\bibitem{Acacio} J. Acacio de Barros   et al,
 Causal Interpretation of Spherically Symmetric Evaporating Black Holes, (2003)\\ www.sbf1.if.usp.br/eventos/enfpc/xxiv/
\bibitem{Takakura}  G. Oliveira-Neto  and F. I. Takakura, Formation of Schwarzschild Black Hole From the Scalar Field Collapse in Four Dimensions, (2003)\\
 www.sbf1.if.usp.br/eventos/enfpc/xxiv/ 
\bibitem{Rufini} R.  Rufini et al,   New perspectives in Physics and Astrophysics from the Theoretical Understanding of Gamma-Ray Bursts, 
 "Proceedings of the $X^{th}$ Brazilian School of Cosmology and Gravitation", M. Novello, editor, AIP, in press, astro-ph/0302557, 
\bibitem{Piran} T. Piran, Gamma-ray Bursts: A Primer  for  Relativists,   gr-qc/0205045
\bibitem{Opher} R. Opher, Can Black Holes Supply  the Energy Needed for Quasars and  Gamma Gay Bursts?,  (2003)\\    www.sbf1.if.usp.br/eventos/enfpc/xxiv/ 
\bibitem{Joras} S. Joras  and  F. D. Sasse, Scalar  Wavepacket Superradiance, (2003)\\ 
 www.sbf1.if.usp.br/eventos/enfpc/xxiv/ 
\bibitem{Fagundes} H.V. Fagundes, On The Visualization of Bolyai-Lobatchevsky's Geometry,   IFT P 03/78-S\~ao Paulo, (1978). Ibid,  The  Intuition of Curved Spaces.
IFT-P 12/80-S\~ao  Paulo, (1980)




\end{thebibliography}
\end{document}